\documentclass{emulateapj} 
\usepackage{rotating}

\begin{document}
\slugcomment{Crepp et al. 2009, ApJ, 702, 672} \thispagestyle{empty}

\title{Direct Detection of Planets Orbiting Large
Angular Diameter Stars: Sensitivity of an Internally-Occulting
Space-based Coronagraph}

\author{Justin R. Crepp\altaffilmark{1}, Suvrath Mahadevan\altaffilmark{2}, Jian Ge}
\affil{Department of Astronomy \\
University of Florida \\
211 Bryant Space Science Center, Gainesville, FL 32611}
\email{jcrepp@astro.caltech.edu}

\altaffiltext{1}{Current Address: Department of Astronomy,
California Institute of Technology, 1200 E. California Blvd.
Pasadena, CA 91125} \altaffiltext{2}{Current Address: Department of
Astronomy \& Astrophysics, Pennsylvania State University, 525 Davey
Lab, University Park, PA 16802}

\begin{abstract}
High-contrast imaging observations of large angular diameter stars
enable complementary science questions to be addressed compared to
the baseline goals of proposed missions like the Terrestrial Planet
Finder-Coronagraph, New World's Observer, and others. Such targets
however present a practical problem in that finite stellar size
results in unwanted starlight reaching the detector, which degrades
contrast. In this paper, we quantify the sensitivity, in terms of
contrast, of an internally-occulting, space-based coronagraph as a
function of stellar angular diameter, from unresolved dwarfs to the
largest evolved stars. Our calculations show that an assortment of
band-limited image masks can accommodate a diverse set of
observations to help maximize mission scientific return. We discuss
two applications based on the results: the spectro-photometric study
of planets already discovered with the radial velocity technique to
orbit evolved stars, which we elucidate with the example of Pollux
b, and the direct detection of planets orbiting our closest
neighbor, $\alpha$ Centauri, whose primary component is on the
main-sequence but subtends an appreciable angle on the sky. It is
recommended that similar trade studies be performed with other
promising internal, external, and hybrid occulter designs for
comparison, as there is relevance to a host of interesting topics in
planetary science and related fields.
\end{abstract}


\section{INTRODUCTION}\label{sec:intro}
High-contrast imaging is contingent upon the destructive
interference of starlight. Losses in spatial coherence due to the
finite size of a star, whose surface is comprised of many
independently radiating elements, can limit the imaging sensitivity
of a coronagraph. The effect is important when attempting to
generate contrast levels that exceed several billion in reflected
light.

A number of nearby stars have sizable angular diameters compared to
the spatial resolution of a large optical telescope. Evolved stars,
in particular, have intrinsically large radii and may subtend
several instrument diffraction widths. For example, the largest star
in the sky, R Doradus, is a red giant of diameter $D_{*}=57 \pm 0.5$
mas \citep{bedding_97}. It illuminates an `area of coherence'
\citep{born_wolf_99}, in quasi-monochromatic light centered on
$\bar{\lambda}=0.55 \; \mu$m, that is 0.6m in diameter --
significantly smaller than the primary mirror with which
large-strategic or flag-ship status missions may operate
\citep{beckwith_08}. For comparison, a 1$R_{\odot}$ star located at
10 pcs would have an angular diameter of $D_{*}=0.9$ mas and
coherently illuminate an area of 40.3m in diameter over the same
bandpass.

An interesting regime lies between the spatially unresolved stars,
where the performance of coronagraphs is nominally quoted, and large
angular diameter giants, where performance is severely affected.
This intermediate observational parameter space contains several
dozen luminosity class I-IV stars and $\alpha$ Centauri A, whose
proximity makes for an exceptional case. Such stars enable
complementary science questions to be addressed compared to the
goals of missions like the proposed Terrestrial Planet Finder --
Coronagraph (TPF-C; \cite{traub_07}), New World's Observer (NWO;
\cite{cash_06}), Advanced Technology Large-Aperture Space Telescope
(ATLAST; \cite{stapelfeldt_09}), and others, whose primary objective
is to search for ``Earth-like" planets orbiting F, G, K
main-sequence stars.

For example, evolved stars can offer access to a different range in
stellar mass and luminosity and thus opportunities for testing
theories of planet formation and evolution in a variety of
environments. Radial velocity surveys have recently begun to explore
this territory. Their preliminary studies of the Jovian planet
population around intermediate-mass stars have already revealed
several apparent trends, including relationships between stellar
mass and planet mass and period \citep{johnson_07,johnson_08}.
Several of these planets are directly detectable and warrant imaging
confirmation and further characterization. One of the planets orbits
a K0 III star, Pollux, which has an angular diameter of 8.0 mas.

The $\alpha$ Centauri system is located 1.4 pcs away from the Sun.
Its primary components are the two highest priority targets on the
TPF-C Top 100 list,\footnote{http://planetquest.jpl.nasa.gov/TPF-C/}
but they have angular diameters of $8.6$ and $6.0$ mas. Special
measures must be taken to observe such stars with the aforementioned
coronagraphic missions, whose primary mirrors range from
$D_{\mbox{tel}}=4-16$m in diameter, since a significant amount of
starlight can leak around the coronagraphic mask and reach the
detector before the stellar surface is fully resolved ($D_* <
\lambda_{\mbox{min}} / D_{\mbox{tel}}$) -- an effect that we
quantify in this paper.


To image a planet orbiting an extended source, an external occulter
will need to cover (occult) more solid-angle on the sky compared to
that of a ``point"-source. This may be accomplished by decreasing
the distance between the star-shade and telescope \citep{lo_AAS09}.
An internal occulter can switch masks in a focal plane wheel, but
resolved patches of the stellar surface will fall onto different
mask locations and create non-common-path errors that can only be
partially corrected with a deformable mirror. Since there exist
trade-offs between sensitivity, inner-working-angle, operating mode,
duty-cycle efficiency, source brightness, signal-to-noise ratio,
cadence, and other parameters, observational plans must account for
the effect of stellar size to balance mission goals with mission
lifetime. The fraction of time that a particular mission will
dedicate to observations of large angular diameter stars is
currently unclear. However, the $\alpha$ Centauri system places a
firm non-zero lower limit.

While knowledge of instrument sensitivity as a function of stellar
size is pertinent for prioritizing targets, especially if the
observing strategy aims to be flexible and incorporate recent
discoveries, the literature is currently devoid of such information
for stars with large angular diameters. In this paper, we calculate
the contrast that an internally occulting, space-based coronagraph
can generate as a function of stellar angular diameter for different
operating modes. The particular design we model utilizes
fourth-order and eighth-order band-limited image masks
\citep{kuchner_traub_02,kcg_05}.

The effects of low-order optical aberrations on band-limited Lyot
coronagraphs have been studied with analytic and numerical
techniques \citep{sg_05,siv_lloyd_05,crepp_07} and lab experiments
\citep{crepp_06}. Here, we present a full treatment of spatial
incoherence that arises from resolving the stellar surface. Our
end-to-end simulation tools explicitly model the dominant physical
effects and incorporate the most recent wavefront correction
algorithm: `electric-field conjugation' \citep{giveon_07}.

Since large angular diameter stars represent an important family of
high-contrast imaging targets, we present these results in
anticipation of comparing them to other promising coronagraphic
designs, like PIAA \citep{guyon_03}, distant and not-so-distant
external occulters \citep{cash_06,spergel_09}, interferometric
coronagraphs \citep{shao_09}, shaped pupils \citep{kasdin_03},
apodized pupils \citep{soummer_05}, vortex masks \citep{mawet_07},
and others; each can, in theory, compensate for stellar size.
\cite{guyon_06} have compared these designs on an equal footing
using a `useful throughput' metric, but only for sources with an
extent of $0.2 \; \lambda / D_{\mbox{\mbox{tel}}}$. Also,
information on wavefront quality and correction capabilities must be
supplied to convert useful throughput into contrast. In the
following, we calculate the intensity of residual starlight within
the dark-hole search area over the entire range of stellar angular
diameters, from unresolved dwarfs to the largest evolved stars.
These numerical simulations can be used to facilitate feasibility
studies for current and future science cases. Based on the results,
we briefly discuss two applications within the context of extrasolar
planet imaging.

\section{NUMERICAL SIMULATIONS}
\label{sec:technique}

We model a TPF-C-like instrument with code written in Matlab
assuming an internally occulting Lyot-style design. Diffraction is
managed with band-limited image masks
\citep{kuchner_traub_02,kcg_05}. The telescope is circular and
unobscured. Simulations are broadband and incorporate primary mirror
phase and amplitude errors, image mask phase errors, a single,
noiseless 64x64 actuator deformable mirror (DM), and the finite size
of the star.

Stars are modeled with a uniform disk of mutually incoherent point
sources. The effects of limb-darkening are also explored in a single
case. Light from each source is sent through the optical train with
a tip/tilt error that corresponds to its location on the stellar
surface. The number of sources across the disk well exceeds the
Nyquist frequency in $\lambda_{\mbox{min}}/D_{\mbox{tel}}$ units.
The intensities from each add together at the detector to form the
final image. Ten wavelength channels sample each of three different
100 nm wide bandpasses. All optics are located in a pupil or image
plane and Fourier transforms are used to propagate the
electric-field. We refer the reader to \cite{sg_06} for a discussion
of Fresnel propagation effects, such as phase-induced amplitude
aberrations and amplitude-induced phase-aberrations, which can limit
a coronagraph's broadband performance. They derive optical quality
requirements for a contrast level of $10^{-12}$ per
spatial-frequency at visible wavelengths over the same bandpass --
an order of magnitude fainter than the scattered light levels dealt
with in this study.

Primary mirror phase errors follow a broken power-law
power-spectral-density (PSD) given by:
\begin{equation}
\mbox{PSD}(k)=\frac{A_0}{1+(k / k_0)^n} \label{equ:psd}
\end{equation}
where $A_0 = 9.6\times10^{-19} \; \mbox{m}^4$, $k$ is the spatial
frequency, $k_0=4$ cycles / m, and $n=3$. This is the PSD typical of
an 8m primary mirror \citep{sg_06,borde_traub_06}. The mirror
surface figure is scaled to have an rms value of 1 nm (2 nm in
wavefront phase). Amplitude errors are modeled as white noise with
an rms of $0.005$ and maximum value of unity. We do not model the
effects of polarization-induced low-order aberrations
\citep{elias_04,bo_04}, which can be suppressed to contrast levels
below $10^{-10}$ for point sources with an eighth-order mask
\citep{sg_05}, nor scattering from other optics in the path other
than the glass on the image mask.

Mask defects, such as phase-shifts that result from the
manufacturing process, can limit the achievable contrast in
polychromatic light for point sources
\citep{lay_05,bala_08,moody_08}. Resolved sources add an additional
complication in that different sections of the star's surface fall
onto different mask locations and create non-common-path errors that
cannot each be compensated for simultaneously; correcting one error
may amplify another. We incorporate these effects into our
simulations by modeling mask imperfections as phase errors that are
correlated on a scale of $0.25 \; \lambda_{\mbox{min}} /
D_{\mbox{tel}}$.


We use both 4th-order and 8th-order linear band-limited masks to
make a comparison study since they have a different resistance to
stellar size. Their amplitude transmissions follow
$\mbox{sinc}(..)^2$ and $\mbox{sinc}(..)+\mbox{sinc}^2(..)$ profiles
respectively (see \cite{kcg_05} for details). The default
inner-working angle is $4 \; \lambda_{\mbox{max}} / D_{\mbox{tel}}$.

The DM is placed at a pupil and its surface is shaped by a square
grid of actuators that map perfectly onto the primary mirror. The
influence of each actuator is modeled with a Gaussian function that
drops to 6\% of its peak value at the location of adjacent
actuators. The 64x64 system can correct for wavefront errors with
spatial frequencies as high as 32 cycles per aperture. This creates
the familiar ``dark-hole" region in the image plane that defines the
search area \citep{trauger_04,trauger_traub_07}. We sacrifice
correction of the highest spatial frequencies, 30-32 cycles per
aperture, to improve correction of lower-order modes, further
reducing the intensity of speckles close to the optical axis. A
smaller search area can yield even deeper contrast, but here we are
also interested in distant planets.


We calculate the optimal DM shape using the `electric-field
conjugation' algorithm developed by \cite{giveon_07}. The technique
is quite general and produces deeper dark-holes compared to previous
techniques, like speckle-nulling, by a factor of order unity.
Successful implementation relies upon accurate modeling of the
coronagraph, electric-field reconstruction at the science camera,
and a form of phase diversity to solve for the actuator heights. The
procedure is efficient once a rather computationally expensive
metric, the ``G-matrix", is established for the optical system. It
needs calculating only once, unless changes to the coronagraph are
made. In this paper, we use several different coronagraphs, of
different mask order, bandpass, inner-working angle, and size of
mask phase errors, so it was necessary to utilize multiple
processors in parallel (but with no message passing interface).

The optimal DM shape is found using only the central (on-axis)
portion of the star. The surface is then fixed -- a DM cannot
compensate for stellar size -- and contrast is measured as a
function of angular diameter. In practice, wavefront sensing may be
performed on a main-sequence star with small angular diameter; then
the telescope can slew to the target of interest.


Results use information from a single image. The instantaneous
contrast, $C(x,y)$, is found using the formula in
\cite{green_shaklan_03} and \cite{crepp_07}, which we show here:
\begin{equation}\label{equ:con}
C(x,y)=\frac{I(x,y)}{I(0,0) \: |M(x,y)|^2},
\end{equation}
where $I(x,y)$ is the intensity at the coordinates ($x$, $y$) in the
final image, $I(0,0)$ is the peak stellar intensity that would be
measured without the image mask in the optical train, and
$|M(x,y)|^2$ is the mask intensity transmission. Both $I(x,y)$ and
$I(0,0)$ are measured with the Lyot stop in place. Linear masks have
no dependence on $y$. Below, we evaluate contrast by taking the
median value of $C(x,y)$ in a $\lambda_{\mbox{min}} /
D_{\mbox{tel}}$ wide box whose inner-edge is located at the
inner-working angle in the most sensitive half of the dark-hole.


\section{CONTRAST VS. ANGULAR DIAMETER}
\label{sec:results}

We compare 4th and 8th-order masks in systems optimized for two
different bandpasses at visible wavelengths, $\lambda=0.5-0.6 \;
\mu$m and $\lambda=0.7-0.8 \; \mu$m, and one in the near-infrared,
$\lambda=2.2-2.3 \; \mu$m. The latter bandpass is included since
planets may exhibit interesting spectral features in the
near-infrared \citep[and references
therein]{kaltenegger_et_al_07,beckwith_08} and there are several
practical benefits regarding dark-hole depth and width as the
wavelength increases. We also include a mask with a larger
inner-working-angle (hereafter, IWA) and place limits on the size of
mask phase errors as a function of stellar diameter.

\begin{figure*}[!ht]
\centering
\includegraphics[height=3.5in]{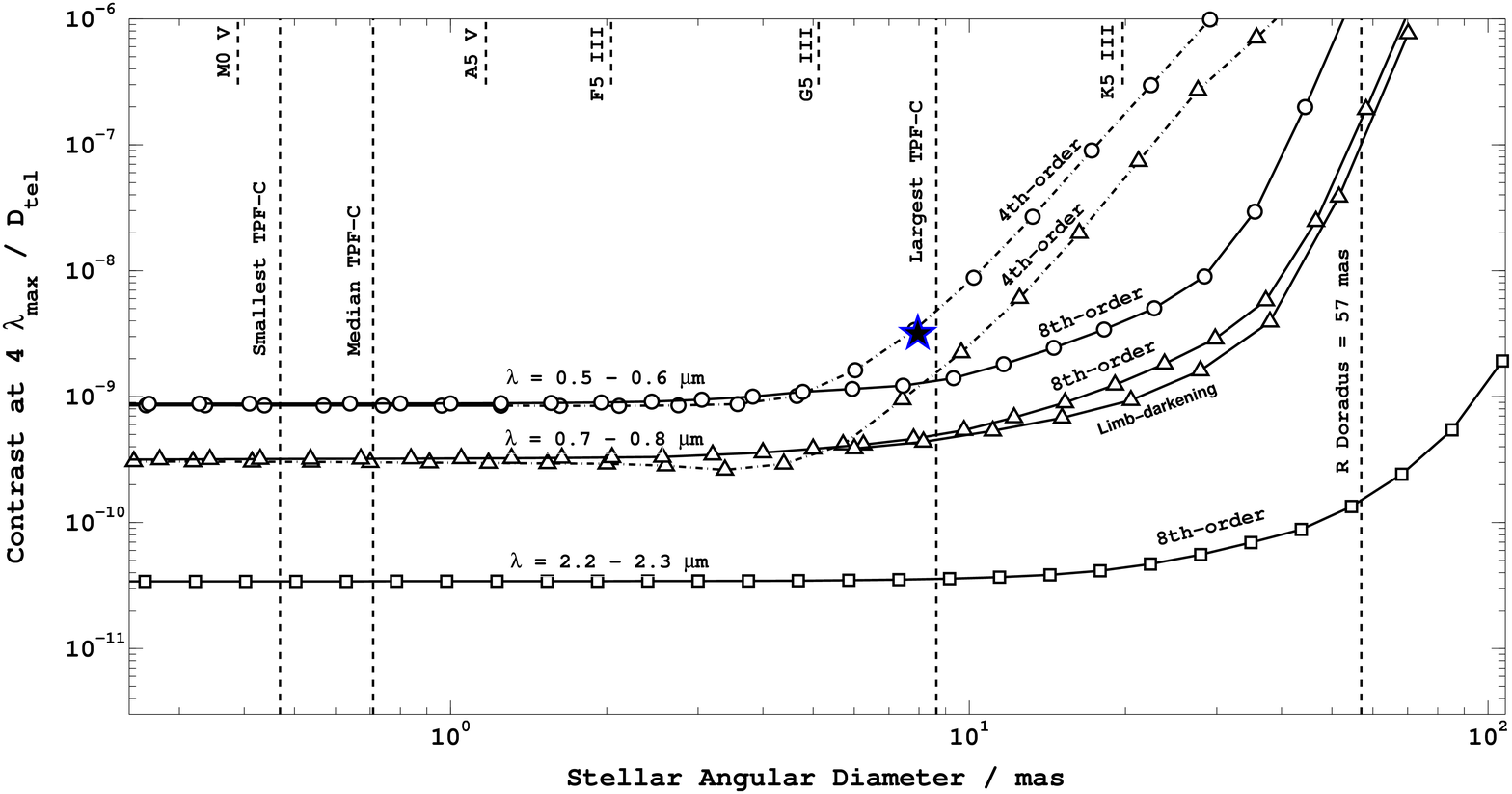} \\
\vspace{1.2cm}
\includegraphics[height=3.5in]{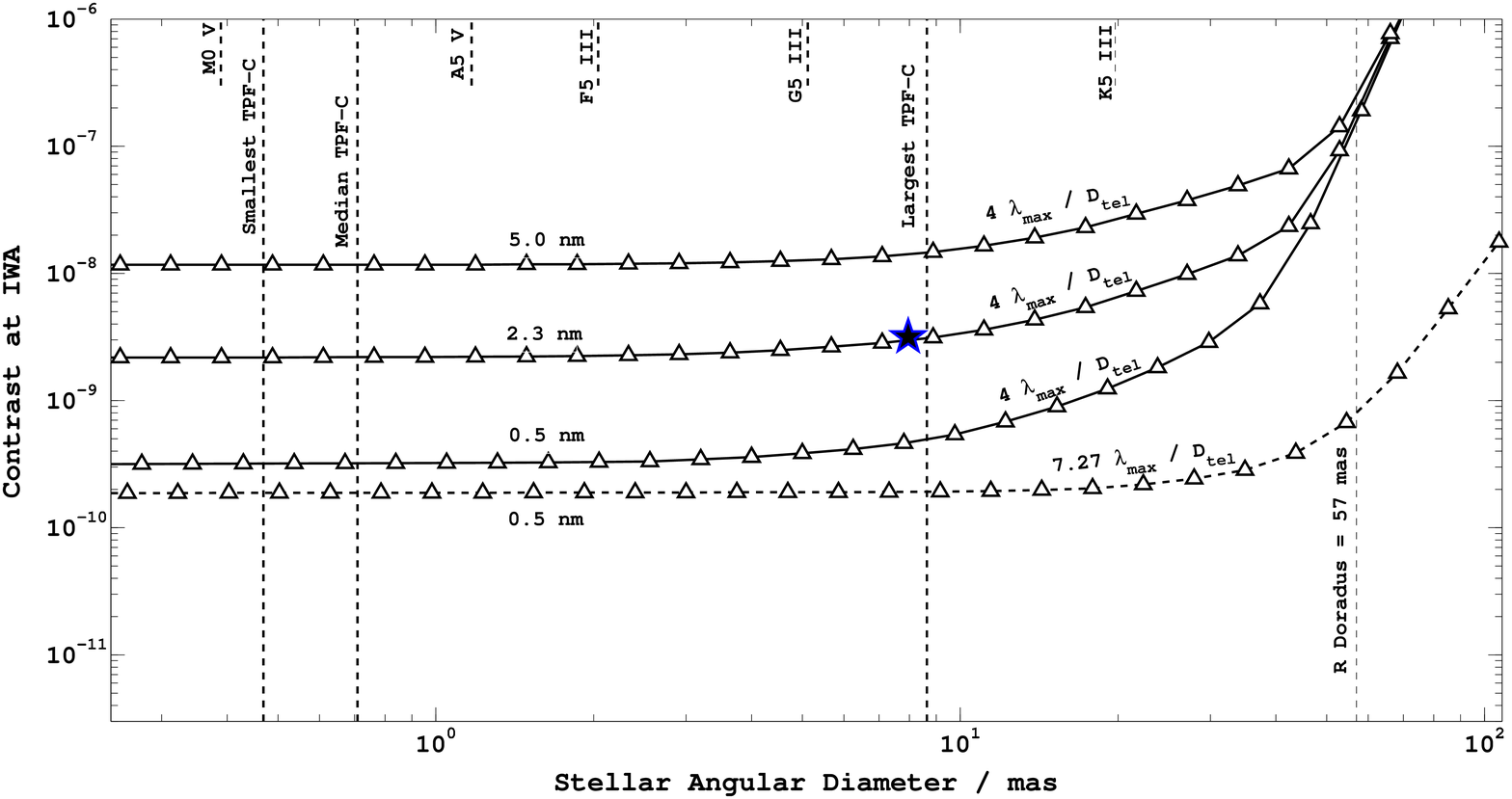}
\caption{\scriptsize Coronagraph sensitivity to stellar angular
diameter for an 8m telescope. The characteristic size of various
stars placed at 15 pcs are shown across the top axis in each graph
for reference. We also include the diameter of R Doradus, the
largest star in the sky, and entries from the TPF-C Top 100 List. A
filled star denotes the approximate reflected light contrast of the
Jovian planet orbiting Pollux. (Top) Fourth-order and eighth-order
mask performance in three different bandpasses. Different data
points (circles, triangles, squares) denote different bandpasses.
The size of mask phase errors is 0.5 nm. The largest TPF-C star also
has the highest priority (see $\S$\ref{sec:alpha_cen} on $\alpha$
Centauri A). Limb-darkening is considered for the $\lambda=0.7-0.8
\;\mu$m, eighth-order mask case. (Bottom) Performance of an
eighth-order mask with different size mask phase errors and a design
with an IWA of $7.27 \; \lambda_{\mbox{max}}/D_{\mbox{tel}}$, each
optimized for the $\lambda=0.7-0.8 \;\mu$m bandpass.}
\label{fig:con_vs_D}
\end{figure*}

Results from our simulations are shown in Fig.~\ref{fig:con_vs_D}
where we plot contrast versus angular diameter for stars wider than
0.2 mas. The smallest, median, and largest targets from the TPF-C
Top 100 list and the largest star in the sky, R Doradus, are shown
for comparison. The upper horizontal axis of each plot also
indicates the characteristic diameter of a variety of stars placed
at 15 pcs. A filled star denotes the approximate location of Pollux
b in reflected light ($\S$\ref{sec:mass_age}).

The top panel of Fig.~\ref{fig:con_vs_D} compares 4th-order and
8th-order masks in different bandpasses. Each has an IWA of $4 \;
\lambda_{\mbox{max}} / D_{\mbox{tel}}$. In the small star regime,
where $D_* << \lambda_{\mbox{min}} / D_{\mbox{tel}}$, contrast is
limited by DM fitting errors, and the sensitivity to stellar size is
constant over most of the TPF-C target list. Further increases to
$D_*$, however, show that all masks leak starlight before the
stellar surfaces are fully resolved:
$\lambda_{\mbox{min}}/D_{\mbox{tel}}=14.2$, $19.3$, and $58.0$ mas
for each bandpass respectively. In this regime, where $D_* \sim
\lambda_{\mbox{min}} / D_{\mbox{tel}}$, the 8th-order mask has a
significantly higher tolerance to stellar size than the 4th-order
mask. This result makes intuitive sense since the 8th-order mask is
also less susceptible to tip/tilt errors
\citep{kcg_05,sg_05,crepp_06,crepp_07}. The difference between the
effects, however, is twofold: (i) a star delivers a collection of
tip/tilt errors to the coronagraph, each with a contribution to the
diffracted light noise floor that depends on distance from the
optical axis, and (ii) the speckle halo from each source on the
stellar surface will couple differently to mask errors. Our
simulations ($\S\ref{sec:technique}$) are able to capture such
details.

We find that an 8th-order mask is required to optimize sensitivity
to the evolved star Pollux, which is known to host a long period
planet, and the largest TPF-C star, $\alpha$ Cen A, which has a high
target priority (see $\S\ref{sec:applications}$). In the extreme
case of R Doradus, instantaneous contrast levels of order $10^{-7}$
and $10^{-10}$ at 4 $\lambda_{\mbox{max}} / D_{\mbox{tel}}$ are
possible with an 8th-order mask for visible and near-infrared
applications respectively.

For small stars, the contrast improves at longer wavelengths
according to the scaling relation $C \propto (\lambda_0/\lambda)^2$,
where $\lambda_0$ and $\lambda$ are the central wavelength of the
original and current bandpass respectively. The sensitivity in the
near-infrared is an order of magnitude deeper than in the visible,
at the expense of a factor of 3-4 in spatial resolution. This
trade-off is well justified for planets with large orbital
separations ($\S$\ref{sec:mass_age}). The relative improvement grows
to several orders of magnitude with further increases in stellar
angular diameter. Ground-based instruments will be able to probe the
$C\geq10^{-7}$ in the near-infrared over all stellar sizes with
extreme adaptive optics
\citep{macintosh_GPI_06,dekany_07,beuzit_06}.

Limb-darkening becomes important when the stellar surface is
resolved. In this regime, the contrast improves by a factor of order
unity compared to that of a uniform disk, since the outer-most
portions of the star -- those that contribute most to the noise
floor -- are less intense \citep{wade_85}. The second
$\lambda=0.7-0.8 \; \mu$m, eighth-order mask contrast curve assumes
a radial intensity profile that is gray (the same for all
wavelengths in the band) and decreases linearly away from the star
center-point. The relative intensity at the edge of the star was set
to 0.6, the value determined by a recent interferometric study of
the red giant star $\epsilon$ Oph \citep{mazumdar_09}.
Limb-darkening coefficients are smaller in the near-infrared
\citep{claret_00} and thus the effects less pronounced.

The bottom panel of Fig.~\ref{fig:con_vs_D} compares 8th-order masks
with rms surface errors of 0.5, 2.3, and 5.0 nm. An 8th-order mask
with an IWA of 7.27$\;\lambda_{\mbox{max}} / D_{\mbox{tel}}=150$ mas
is also shown. Each was optimized for the $0.7-0.8 \;\mu$m bandpass.
The rms=0.5 nm, IWA=4$\;\lambda_{\mbox{max}}/D_{\mbox{tel}}$ curve
is the same from the upper panel. Increasing the size of mask errors
scatters more light into the dark hole. For unresolved stars, the
contrast scales according to $C\propto(\sigma/\sigma_0)^2$, where
$\sigma_0$ and $\sigma$ are the total rms error from the mask and
primary mirror added in quadrature for two different cases.
Presumably, other uncorrelated errors, such as polarization effects,
may also be added in quadrature to estimate how the scattered light
noise floor may change. In this sense, Fig. 1 serves as a
calibration point for calculating the contrast of systems with
different error budgets. The coronagraphic mask with IWA$=
7.27\;\lambda_{\mbox{max}}/D_{\mbox{tel}}$ passes more planet light
and is also less susceptible to stellar size (see
$\S\ref{sec:mass_age}$).


\begin{figure*}[!ht]
\centering
\includegraphics[width=6.4in]{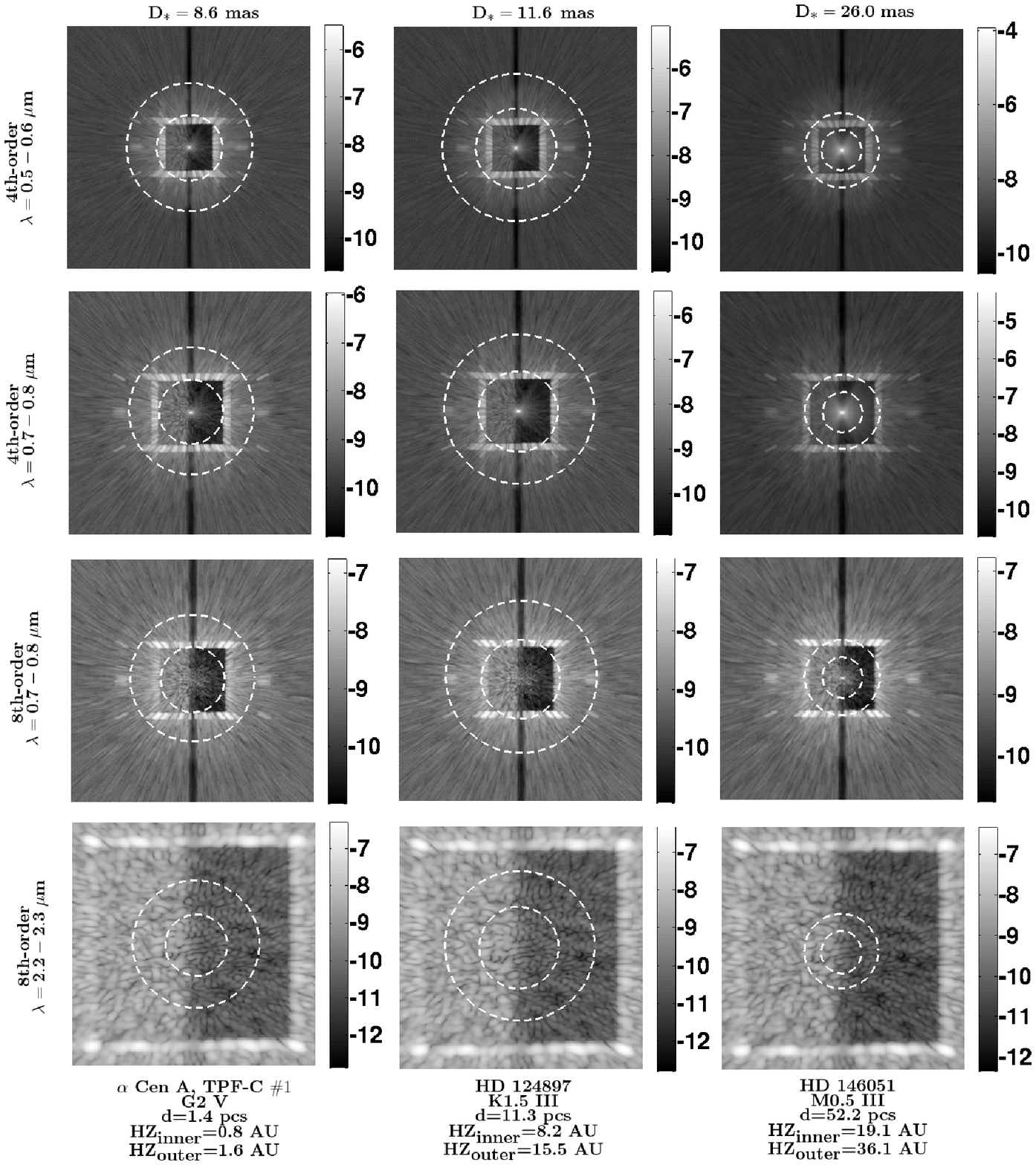}
\caption{\scriptsize High-contrast images of three stars with large angular
diameters. Each are more massive than the Sun. The gray-scale shows
the value of $\mbox{log}(I(x,y)/I(0,0))$. The dashed white lines
indicate the location of the habitable zone for reference. Amplitude
errors break the spatial symmetry of speckles allowing optimal
suppression of only half of the dark-field with a single deformable
mirror. The 4th-order mask leaks a significant amount of starlight.
Observations in the near-infrared provide a wider dark-hole and
deeper contrast, thus permitting the reflected light detection of
more distant planets.} \label{fig:images}
\end{figure*}

In practice, an internal occulter will need to implement
post-processing techniques after creating a dark-hole. However, the
improvement in sensitivity of a coronagraph that utilizes
differential-imaging is nearly constant as a function of stellar
size. We have verified this assertion with simulations of PSF
roll-subtraction for stars as large as R Doradus and telescope
diameters as large as 8m. Figure~\ref{fig:con_vs_D} can thus also be
used to estimate the planet-imaging sensitivity of an instrument
with speckle-discrimination capabilities.

The effect of stellar size will become increasingly problematic with
time as telescope diameters and interferometric baselines continue
to grow. To first order, we can scale the curves in
Fig.~\ref{fig:con_vs_D} according to telescope size, if we assume a
constant actuator density and neglect differences in the
manufacturing processes that modify the PSD and thus the speckle
pattern. For instance, a larger telescope with a correspondingly
larger number of actuators would better resolve the stellar surface,
leak more light, and result in a horizontal curve shift (to the
left) by a factor of $8 / D_{\mbox{tel}} \mbox{(m)}$. Telescope
designs with primary mirror diameters in excess of 8m, which may
eventually be required to search a statistically significant sample
of stars for terrestrial planets in reflected light
\citep{beckwith_08}, are thus presented with a dilemma.


Figure~\ref{fig:images} displays a series of high-contrast images
for three nearby stars of increasing angular diameter. The current
inner and outer-edge of the habitable zone for each is overlaid
(dashed white lines) for reference. Only half of the dark-hole is
accessible since we include primary mirror amplitude errors and a
single deformable mirror. The gray-scale indicates the value of
$\mbox{log}(I(x,y)/I(0,0))$. The images qualitatively verify the
results from Fig.~\ref{fig:con_vs_D}. Stellar leakage is a serious
problem for the 4th-order mask in both visible bandpasses, whereas
the 8th-order mask is able to suppress starlight from extended
sources to more acceptable levels. Sacrificing spatial resolution by
conducting observations in the near-infrared improves the
coronagraph's resistance to stellar size and dark-hole depth and
width. Equipping a space-based coronagraph with a near-infrared
camera would increase the signal-to-noise ratio of spectra, enable
the detection of planets at larger separations, and complement
visible light observations. We next discuss two applications based
on these results.

\section{POTENTIAL APPLICATIONS}\label{sec:applications}

\subsection{Evolved Stars}\label{sec:mass_age}
Stars supply the material out of which planets originate. One might
therefore expect some diversity in the statistical properties of
planets, given that the range of masses over the
zero-age-main-sequence typically spans three orders of magnitude.
Studies of the planet population as a function of stellar mass may
provide insight into the processes that govern their formation. A
similar argument can be made with respect to the time evolution of
planets. Planetary atmospheres are subject to the radiation they
receive from their host star and stellar luminosities can vary by as
many as seven orders of magnitude over a lifetime, when the
red-giant to early compact object stages of evolution are included
\citep{iben_67,iben_laughlin_89}. Combining this information
suggests that evolved stars may become important targets for planet
searches, if they are not already. Direct imaging can offer an
efficient approach for probing this parameter space, as has been
recently demonstrated by the simultaneous discovery of three planets
orbiting HR 8799 \citep{marois_08}.

A low-risk approach for large-aperture missions to target evolved
stars is to follow-up planets that have already been detected with
other techniques, like radial velocity and astrometry. An example of
such a system with a known exoplanet is the bright (V=1.15) K0 III
giant $\beta$ Gem (Pollux). Radial velocity observations spanning
nearly 25 years are consistent with the presence of a $M
\mbox{sini}=2.6M_{\mbox{Jup}}$ planet with a semi-major axis of 1.64
AU \citep{hatzes_06}. Pollux is located at a distance of 10.3 pcs
and has a mass of $1.7 \pm 0.4 \; M_{\odot}$ \citep{allende_99} and
angular diameter of $7.96 \pm 0.09$ mas \citep{nordgren_01}. The
approximate location of Pollux b is labeled with a star in
Fig.~\ref{fig:con_vs_D}. According to the curves, an 8th-order mask
is required to unambiguously detect its presence at $\lambda =
0.5-0.6$ $\mu$m, assuming half illumination, a planet radius of
$1R_{\mbox{Jup}}$, and albedo of 0.3. The orbits of such planets
will already be determined, and the combination of photometry and
spectra with indirect techniques will place tight constraints on
their mass.


We also note that terrestrial planets can be imaged in the extended
habitable zones of evolved stars. A recent study by \cite{lsd05} has
shown that there are eras of post-main-sequence evolution where
hospitable conditions may persist for $10^8-10^9$ years. Ten percent
of the stars within 30 pcs of the Sun that they recommend for a
targeted search have angular diameters between $3 \lesssim D_{*}
\lesssim 57$ mas. This same set of stars from their sample also have
the brightest visual magnitudes, so may offer high observational
efficiency.


Although distant terrestrial planets will generally be fainter than
the Earth-Sun system in reflected light (contrast $\approx
2\times10^{-10}$), depending on their albedo and radius (see
\cite{seager_07} for a discussion of super-Earths), a coronagraph
can accommodate by sacrificing spatial resolution. Observations at
longer wavelengths decreases the intensity of speckles by a factor
of $\sim(\lambda / \lambda_0)^2$ in the search area
(\S\ref{sec:results}). Another option is to increase the
inner-working-angle (Fig.~\ref{fig:con_vs_D} -- bottom panel). With
a band-limited mask design, this results in higher Lyot stop
throughput, which increases the amount of companion light, decreases
integration time, and makes the point-spread-function (PSF) more
spatially succinct.\footnote{Highly concentrated PSF's facilitate
discrimination of companion light amongst diffuse zodiacal and
exozodiacal dust emission.} Notice that both approaches, and
combinations thereof, improve the coronagraph's resistance to
stellar size.




\subsection{$\alpha$ Centauri}\label{sec:alpha_cen}
The primary components of the $\alpha$ Centauri (Cen) triple star
system are on the main-sequence but have angular diameters of 8.6
and 6.0 mas respectively (Table~\ref{tab:alpha_cen}). They are the
two highest priority targets on the TPF-C Top 100 List, where
priority is defined by the first-visit completeness per unit
integration time. The third component, Proxima Centauri, is a
distant M-dwarf that appears to be weakly bound
\citep{wertheimer_06}.

Numerical simulations show that terrestrial planets located in the
habitable zone of $\alpha$ Cen A, B may remain stable for timescales
comparable to the system age, provided the inclination angle between
the stellar and planetary orbital planes is sufficiently small
\citep{wiegert_1997,quintana_07}. The topic of whether terrestrial
planets can form in their habitable zone in the first place though
is still debated \citep{barbieri_02,thebault_08a,thebault_08b}.
Imaging observations of intermediately spaced binaries can provide
important data to further the development of planet formation
models.

A vertical dashed line denotes the location of $\alpha$ Cen A in
Fig.~\ref{fig:con_vs_D} as the largest TPF-C star. We find that an
8th-order mask generates deeper contrast at the IWA than a 4th-order
mask by factors of 3.1 and 3.6 at $\lambda=0.5-0.6 \; \mu$m and
$\lambda=0.7-0.8 \; \mu$m respectively. The first column of
Fig.~\ref{fig:images} shows simulated images of $\alpha$ Cen A in
various observing modes and bandpasses. Given the relationship
between wavelength and dark-hole depth and width
(\S\ref{sec:results}), planets in the habitable zone of $\alpha$ Cen
A may be easier to detect in the near-infrared.

\begin{table}[!th]
\vspace{-0.1in} \centerline{
\begin{tabular}{lccc}
$\alpha$ Centauri    &      A       &     B       &     C            \\
\hline
Spec. Type           &    G2 V      &   K1 V      &    M5.5 V        \\
V mag                &   -0.01      &   1.35      &     11.1         \\
mass  ($M_{\odot}$)  &    1.11      &   0.93      &     0.12         \\
radius ($R_{\odot}$) &    1.24      &   0.87      &     0.15         \\
luminosity ($L_{\odot}$) & 1.60     &   0.45      &    0.0002        \\
$[$Fe/H$]$           &     0.22     &   0.26      &    -1.00         \\
a (AU)               &     23.4     &   23.4      & $\approx$ 12,000 \\
e                    &     0.52     &   0.52      &      ?           \\
i                    &    79.2$^{\mbox{o}}$  &   79.2$^{\mbox{o}}$  & ? \\
diameter (mas)       &     8.6      &   6.0       &     1.0          \\
TPF-C Rank           &    $\#1$     &    $\#2$    &     --          \\
\hline
\end{tabular}}
\caption{Physical parameters of the $\sim 5$ Gyr old $\alpha$ Cen
system.} \label{tab:alpha_cen}
\end{table}



\section{CONCLUDING REMARKS}\label{sec:conclusions}
Our primary objective with this work has been to quantify the
high-contrast imaging sensitivity of an internally occulting,
space-based coronagraph as a function of stellar angular diameter.
Figures~\ref{fig:con_vs_D} and \ref{fig:images} illustrate the
fundamental tradeoffs associated with 4th and 8th-order band-limited
masks in various bandpasses and with different size mask phase
errors and IWA's for all stars in the sky. We have also included a
case that incorporates the effects of limb-darkening. These results
can be used to develop observing strategies and new science cases
for a TPF-C-like instrument, and may be compared to other promising
internal, external, and hybrid occulter designs. We also note that
NIRCAM in JWST will utilize band-limited image masks
\citep{green_05}.

A significant amount of starlight leaks through the coronagraph
before the stellar surface is fully resolved. Consequently, many
important targets will require non-default observational operating
modes to accommodate for stellar size. An 8th-order mask can provide
this leverage by generating as much as two orders of magnitude
deeper contrast than a 4th-order mask for large angular diameter
stars.

We also find that dark-hole depth depends sensitively on the optical
quality of the mask, bandpass, and IWA. Only 5 nm of rms surface
roughness (peak-to-valley $\approx25$ nm) can degrade contrast to
$\sim 10^{-8}$, prior to post-processing techniques. Near-infrared
observations can help to offset the effect by reducing the amount of
scatter from optical components. They also offer a wider dark-hole
and thus may be ideal for detecting the reflected light of distant
planets. There are several dozen targets where this newly afforded
search space corresponds to the habitable zone. Most are evolved
stars. Sacrifices in IWA can offer similar performance at the
expense of search area.



There are two applications that follow immediately from these
results:\footnote{Other applications include stellar multiplicity
studies and the detection and characterization of brown dwarfs and
debris disks.}\vspace{0.1cm}

{\noindent}(a) Jovian planets have been detected orbiting evolved
stars with the radial velocity technique. Their physical properties
may be further constrained with spectro-photometric follow-up. The
planet orbiting Pollux is an example and requires an 8th-order mask
for unambiguous detection. Coronagraphs with intrinsically small
IWA's may not be able to detect Pollux b in reflected light.
Presumably, more such planets will be detected in the future as the
time baselines of radial velocity and astrometric searches
increase.\vspace{0.1cm}

{\noindent}(b) Given its proximity to the Sun, $\alpha$ Centauri
will be a high priority target for any space-based high-contrast
imaging mission (e.g., Kaltenegger et al. 2009). Both of its
main-sequence stars have a large angular diameter. An 8th-order mask
can generate a factor of $\sim 3$ deeper contrast at the IWA than a
4th-order mask for Alpha Centauri A. Near-infrared observations are
likely required to search its habitable zone with deformable mirrors
that utilize 64x64 actuators or less.\vspace{0.1cm}

{\noindent}No other TPF-C stars require an eighth-order mask, unless
a telescope with primary mirror in excess of 8m in diameter is
launched (e.g., ATLAST). Fig.~\ref{fig:con_vs_D} can be used as
calibration for scaling arguments to estimate the expected
performance of different aperture sizes. Small-strategic or
discovery-class missions, like ACCESS \citep{trauger_09}, PECO
\citep{guyon_09}, DAVINCI \citep{shao_09}, EPIC \citep{clampin_09},
or THEIA \citep{spergel_09}, should be able to observe large angular
diameter stars without substantive losses in sensitivity, although
will still require eighth-order masks for a number of targets. The
former application may be a valuable investment of observing time
given that such missions will be limited primarily to the
characterization of Jovian planets, based on arguments of telescope
collecting area and number of stars surveyed \citep{beckwith_08}.
Moreover, it appears that massive stars tend to have more massive
planets at larger separations \citep{johnson_08}.

Given the scientific motivation for observing large angular diameter
stars, it is reasonable that space-based high-contrast imaging
missions consider a more comprehensive set of targets than the
canonical list of F, G, K main-sequence stars. Otherwise, a limited
scope or severe predisposition to certain kinds of stars may neglect
large classes of interesting systems, including those that may tell
us where to look next. Indeed, history has shown that this line of
reasoning also tends to maximize the chances for serendipitous
discovery \citep{angel_01}.

We thank Stuart Shaklan and Amir Give'on for helpful comments
regarding an earlier draft of this manuscript and implementation of
the electric-field conjugation algorithm. We are also grateful for
suggestions made by the referee that lead to improved presentation
of our results. This work was supported by the UCF-UF SRI program
and NASA grant NNG06GC49G.

\bibliographystyle{jtb}
\bibliography{crepp_bib.bbl}

\begin{thebibliography}{}

\bibitem[\protect\astroncite{{Allende-Prieto} and {Lambert}}{1999}]{allende_99}
{Allende-Prieto}, C. and {Lambert}, D.~L. (1999) ,
\newblock {\em \aap} {\bf 352}, 555

\bibitem[\protect\astroncite{{Angel}}{2001}]{angel_01}
{Angel}, R. (2001) ,
\newblock {\em \nat} {\bf 409}, 427

\bibitem[\protect\astroncite{{Balasubramanian}}{2008}]{bala_08}
{Balasubramanian}, K. (2008) ,
\newblock {\em \ao} {\bf 47}, 116

\bibitem[\protect\astroncite{{Barbieri} et~al.}{2002}]{barbieri_02}
{Barbieri}, M., {Marzari}, F., and {Scholl}, H. (2002) ,
\newblock {\em \aap} {\bf 396}, 219

\bibitem[\protect\astroncite{{Beckwith}}{2008}]{beckwith_08}
{Beckwith}, S.~V.~W. (2008) ,
\newblock {\em \apj} {\bf 684}, 1404

\bibitem[\protect\astroncite{{Bedding} et~al.}{1997}]{bedding_97}
{Bedding}, T.~R., {Zijlstra}, A.~A., {von der Luhe}, O., {Robertson}, J.~G.,
  {Marson}, R.~G., {Barton}, J.~R., and {Carter}, B.~S. (1997) ,
\newblock {\em \mnras} {\bf 286}, 957

\bibitem[\protect\astroncite{{Beuzit} et~al.}{2006}]{beuzit_06}
{Beuzit}, J.-L., {Feldt}, M., {Dohlen}, K., {Mouillet}, D., {Puget}, P.,
  {Antichi}, J., {Baruffolo}, A., {Baudoz}, P., {Berton}, A., {Boccaletti}, A.,
  {Carbillet}, M., {Charton}, J., {Claudi}, R., {Downing}, M., {Feautrier}, P.,
  {Fedrigo}, E., {Fusco}, T., {Gratton}, R., {Hubin}, N., {Kasper}, M.,
  {Langlois}, M., {Moutou}, C., {Mugnier}, L., {Pragt}, J., {Rabou}, P.,
  {Saisse}, M., {Schmid}, H.~M., {Stadler}, E., {Turrato}, M., {Udry}, S.,
  {Waters}, R., and {Wildi}, F. (2006) ,
\newblock {\em The Messenger} {\bf 125}, 29

\bibitem[\protect\astroncite{{Bord{\'e}} and {Traub}}{2006}]{borde_traub_06}
{Bord{\'e}}, P.~J. and {Traub}, W.~A. (2006) ,
\newblock {\em \apj} {\bf 638}, 488

\bibitem[\protect\astroncite{{Born} and {Wolf}}{1999}]{born_wolf_99}
{Born}, M. and {Wolf}, E. (1999) ,
\newblock {\em Principles of Optics - 7th expanded ed.}

\bibitem[\protect\astroncite{{Breckinridge} and {Oppenheimer}}{2004}]{bo_04}
{Breckinridge}, J.~B. and {Oppenheimer}, B.~R. (2004) ,
\newblock {\em \apj} {\bf 600}, 1091

\bibitem[\protect\astroncite{{Cash}}{2006}]{cash_06}
{Cash}, W. (2006) ,
\newblock {\em \nat} {\bf 442}, 51

\bibitem[\protect\astroncite{{Clampin}}{2009}]{clampin_09}
{Clampin}, M. (2009) ,
\newblock In {\em American Astronomical Society Meeting Abstracts}, Vol. 213 of
  {\em American Astronomical Society Meeting Abstracts}, p. 234.06

\bibitem[\protect\astroncite{{Claret}}{2000}]{claret_00}
{Claret}, A. (2000) ,
\newblock {\em \aap} {\bf 363}, 1081

\bibitem[\protect\astroncite{{Crepp} et~al.}{2006}]{crepp_06}
{Crepp}, J.~R., {Ge}, J., {Vanden Heuvel}, A.~D., {Miller}, S.~P., and
  {Kuchner}, M.~J. (2006) ,
\newblock {\em \apj} {\bf 646}, 1252

\bibitem[\protect\astroncite{{Crepp} et~al.}{2007}]{crepp_07}
{Crepp}, J.~R., {Vanden Heuvel}, A.~D., and {Ge}, J. (2007) ,
\newblock {\em \apj} {\bf 661}, 1323

\bibitem[\protect\astroncite{{Dekany} et~al.}{2007}]{dekany_07}
{Dekany}, R., {Bouchez}, A., {Roberts}, J., {Troy}, M., {Kibblewhite}, E.,
  {Oppenheimer}, B., {Moore}, A., {Shelton}, C., {Smith}, R., {Trinh}, T., and
  {Velur}, V. (2007) ,
\newblock In {\em Advanced Maui Optical and Space Surveillance Technologies
  Conference}

\bibitem[\protect\astroncite{{Elias} et~al.}{2004}]{elias_04}
{Elias}, II, N.~M., {Bates}, R., and {Turner-Valle}, J. (2004) ,
\newblock In {\em Society of Photo-Optical Instrumentation Engineers (SPIE)
  Conference Series}.  (R.~B. {Hoover}, G.~V. {Levin}, and A.~Y. {Rozanov}
  eds.), Vol. 5555 of {\em Presented at the Society of Photo-Optical
  Instrumentation Engineers (SPIE) Conference}, pp. 248--257

\bibitem[\protect\astroncite{{Give'on} et~al.}{2007}]{giveon_07}
{Give'on}, A., {Belikov}, R., {Shaklan}, S., and {Kasdin}, J. (2007) ,
\newblock {\em Optics Express} {\bf 15}, 12338

\bibitem[\protect\astroncite{{Green} et~al.}{2005}]{green_05}
{Green}, J.~J., {Beichman}, C., {Basinger}, S.~A., {Horner}, S., {Meyer}, M.,
  {Redding}, D.~C., {Rieke}, M., and {Trauger}, J.~T. (2005) ,
\newblock In {\em Techniques and Instrumentation for Detection of Exoplanets
  II. Edited by Coulter, Daniel R., Proceedings of the SPIE, Volume 5905, pp.
  185-195 (2005).}

\bibitem[\protect\astroncite{{Green} and {Shaklan}}{2003}]{green_shaklan_03}
{Green}, J.~J. and {Shaklan}, S.~B. (2003) ,
\newblock In {\em Techniques and Instrumentation for Detection of Exoplanets.
  Edited by Coulter, Daniel R. Proceedings of the SPIE, Volume 5170, pp. 25-37
  (2003).}

\bibitem[\protect\astroncite{{Guyon}}{2003}]{guyon_03}
{Guyon}, O. (2003) ,
\newblock {\em \aap} {\bf 404}, 379

\bibitem[\protect\astroncite{{Guyon} et~al.}{2006}]{guyon_06}
{Guyon}, O., {Pluzhnik}, E.~A., {Kuchner}, M.~J., {Collins}, B., and {Ridgway},
  S.~T. (2006) ,
\newblock {\em \apjs} {\bf 167}, 81

\bibitem[\protect\astroncite{{Guyon et al.}}{2009}]{guyon_09}
{Guyon et al.}, O. (2009) ,
\newblock In {\em American Astronomical Society Meeting Abstracts}, Vol. 213 of
  {\em American Astronomical Society Meeting Abstracts}, p. 234.08

\bibitem[\protect\astroncite{{Hatzes} et~al.}{2006}]{hatzes_06}
{Hatzes}, A.~P., {Cochran}, W.~D., {Endl}, M., {Guenther}, E.~W., {Saar},
  S.~H., {Walker}, G.~A.~H., {Yang}, S., {Hartmann}, M., {Esposito}, M.,
  {Paulson}, D.~B., and {D{\"o}llinger}, M.~P. (2006) ,
\newblock {\em \aap} {\bf 457}, 335

\bibitem[\protect\astroncite{{Iben}}{1967}]{iben_67}
{Iben}, I.~J. (1967) ,
\newblock {\em \araa} {\bf 5}, 571

\bibitem[\protect\astroncite{{Iben} and {Laughlin}}{1989}]{iben_laughlin_89}
{Iben}, I.~J. and {Laughlin}, G. (1989) ,
\newblock {\em \apj} {\bf 341}, 312

\bibitem[\protect\astroncite{{Johnson} et~al.}{2007}]{johnson_07}
{Johnson}, J.~A., {Fischer}, D.~A., {Marcy}, G.~W., {Wright}, J.~T.,
  {Driscoll}, P., {Butler}, R.~P., {Hekker}, S., {Reffert}, S., and {Vogt},
  S.~S. (2007) ,
\newblock {\em \apj} {\bf 665}, 785

\bibitem[\protect\astroncite{{Johnson} et~al.}{2008}]{johnson_08}
{Johnson}, J.~A., {Marcy}, G.~W., {Fischer}, D.~A., {Wright}, J.~T., {Reffert},
  S., {Kregenow}, J.~M., {Williams}, P.~K.~G., and {Peek}, K.~M.~G. (2008) ,
\newblock {\em \apj} {\bf 675}, 784

\bibitem[\protect\astroncite{{Kaltenegger} et~al.}{2007}]{kaltenegger_et_al_07}
{Kaltenegger}, L., {Traub}, W.~A., and {Jucks}, K.~W. (2007) ,
\newblock {\em \apj} {\bf 658}, 598

\bibitem[\protect\astroncite{{Kasdin} et~al.}{2003}]{kasdin_03}
{Kasdin}, N.~J., {Vanderbei}, R.~J., {Spergel}, D.~N., and {Littman}, M.~G.
  (2003) ,
\newblock {\em \apj} {\bf 582}, 1147

\bibitem[\protect\astroncite{{Kuchner} and {Traub}}{2002}]{kuchner_traub_02}
{Kuchner}, M.~J. and {Traub}, W.~A. (2002) ,
\newblock {\em \apj} {\bf 570}, 900

\bibitem[\protect\astroncite{{Kuchner, Crepp,} and {Ge}}{2005}]{kcg_05}
{Kuchner, Crepp,} and {Ge} (2005) ,
\newblock {\em \apj} {\bf 628}, 466

\bibitem[\protect\astroncite{{Lay} et~al.}{2005}]{lay_05}
{Lay}, O.~P., {Green}, J.~J., {Hoppe}, D.~J., and {Shaklan}, S.~B. (2005) ,
\newblock In {\em Techniques and Instrumentation for Detection of Exoplanets
  II. Edited by Coulter, Daniel R. Proceedings of the SPIE, Volume 5905, pp.
  148-161 (2005).}.  (D.~R. {Coulter} ed.), Vol. 5905 of {\em Presented at the
  Society of Photo-Optical Instrumentation Engineers (SPIE) Conference}, pp.
  148--161

\bibitem[\protect\astroncite{{Lo} et~al.}{2009}]{lo_AAS09}
{Lo}, A., {Noecker}, C., {Cash}, W., and {NWO Study Team} (2009) ,
\newblock In {\em American Astronomical Society Meeting Abstracts}, Vol. 213 of
  {\em American Astronomical Society Meeting Abstracts}, p. 404.01

\bibitem[\protect\astroncite{{Lopez} et~al.}{2005}]{lsd05}
{Lopez}, B., {Schneider}, J., and {Danchi}, W.~C. (2005) ,
\newblock {\em \apj} {\bf 627}, 974

\bibitem[\protect\astroncite{{Macintosh} et~al.}{2006}]{macintosh_GPI_06}
{Macintosh}, B., {Graham}, J., {Palmer}, D., {Doyon}, R., {Gavel}, D.,
  {Larkin}, J., {Oppenheimer}, B., {Saddlemyer}, L., {Wallace}, J.~K.,
  {Bauman}, B., {Evans}, J., {Erikson}, D., {Morzinski}, K., {Phillion}, D.,
  {Poyneer}, L., {Sivaramakrishnan}, A., {Soummer}, R., {Thibault}, S., and
  {Veran}, J.-P. (2006) ,
\newblock In {\em Advances in Adaptive Optics II. Edited by Ellerbroek, Brent
  L.; Bonaccini Calia, Domenico. Proceedings of the SPIE, Volume 6272, pp.
  62720L (2006).}, Vol. 6272 of {\em Presented at the Society of Photo-Optical
  Instrumentation Engineers (SPIE) Conference}

\bibitem[\protect\astroncite{{Marois} et~al.}{2008}]{marois_08}
{Marois}, C., {Macintosh}, B., {Barman}, T., {Zuckerman}, B., {Song}, I.,
  {Patience}, J., {Lafreni{\`e}re}, D., and {Doyon}, R. (2008) ,
\newblock {\em Science} {\bf 322}, 1348

\bibitem[\protect\astroncite{{Mawet} et~al.}{2007}]{mawet_07}
{Mawet}, D., {Riaud}, P., {Hanot}, C., {Vandormael}, D., {Loicq}, J.,
  {Baudrand}, J., {Surdej}, J., and {Habraken}, S. (2007) ,
\newblock In {\em Society of Photo-Optical Instrumentation Engineers (SPIE)
  Conference Series}

\bibitem[\protect\astroncite{{Mazumdar} et~al.}{2009}]{mazumdar_09}
{Mazumdar}, A., {Merand}, A., {Demarque}, P., {Kervella}, P., {Barban}, C.,
  {Baudin}, F., {Coude du Foresto}, V., {Farrington}, C., {Goldfinger}, P.~J.,
  {Goupil}, M.~., {Josselin}, E., {Kuschnig}, R., {McAlister}, H.~A.,
  {Matthews}, J., {Ridgway}, S.~T., {Sturmann}, J., {Sturmann}, L., {ten
  Brummelaar}, T.~A., and {Turner}, N. (2009) ,
\newblock {\em ArXiv e-prints}

\bibitem[\protect\astroncite{{Moody} et~al.}{2008}]{moody_08}
{Moody}, D.~C., {Gordon}, B.~L., and {Trauger}, J.~T. (2008) ,
\newblock In {\em Society of Photo-Optical Instrumentation Engineers (SPIE)
  Conference Series}, Vol. 7010 of {\em Presented at the Society of
  Photo-Optical Instrumentation Engineers (SPIE) Conference}

\bibitem[\protect\astroncite{{Nordgren} et~al.}{2001}]{nordgren_01}
{Nordgren}, T.~E., {Sudol}, J.~J., and {Mozurkewich}, D. (2001) ,
\newblock {\em \aj} {\bf 122}, 2707

\bibitem[\protect\astroncite{{Quintana} et~al.}{2007}]{quintana_07}
{Quintana}, E.~V., {Adams}, F.~C., {Lissauer}, J.~J., and {Chambers}, J.~E.
  (2007) ,
\newblock {\em \apj} {\bf 660}, 807

\bibitem[\protect\astroncite{{Seager} et~al.}{2007}]{seager_07}
{Seager}, S., {Kuchner}, M., {Hier-Majumder}, C.~A., and {Militzer}, B. (2007)
  ,
\newblock {\em \apj} {\bf 669}, 1279

\bibitem[\protect\astroncite{{Shaklan} and {Green}}{2005}]{sg_05}
{Shaklan}, S.~B. and {Green}, J.~J. (2005) ,
\newblock {\em \apj} {\bf 628}, 474

\bibitem[\protect\astroncite{{Shaklan} and {Green}}{2006}]{sg_06}
{Shaklan}, S.~B. and {Green}, J.~J. (2006) ,
\newblock {\em \ao} {\bf 45}, 5143

\bibitem[\protect\astroncite{{Shao}}{2009}]{shao_09}
{Shao}, M. (2009) ,
\newblock In {\em American Astronomical Society Meeting Abstracts}, Vol. 213 of
  {\em American Astronomical Society Meeting Abstracts}, p. 234.05

\bibitem[\protect\astroncite{{Sivaramakrishnan} and
  {Lloyd}}{2005}]{siv_lloyd_05}
{Sivaramakrishnan}, A. and {Lloyd}, J.~P. (2005) ,
\newblock {\em \apj} {\bf 633}, 528

\bibitem[\protect\astroncite{{Soummer}}{2005}]{soummer_05}
{Soummer}, R. (2005) ,
\newblock {\em \apjl} {\bf 618}, L161

\bibitem[\protect\astroncite{{Spergel} et~al.}{2009}]{spergel_09}
{Spergel}, D.~N., {Kasdin}, J., {Belikov}, R., {Atcheson}, P., {Beasley}, M.,
  {Calzetti}, D., {Cameron}, B., {Copi}, C., {Desch}, S., {Dressler}, A.,
  {Ebbets}, D., {Egerman}, R., {Fullerton}, A., {Gallagher}, J., {Green}, J.,
  {Guyon}, O., {Heap}, S., {Jansen}, R., {Jenkins}, E., {Kasting}, J.,
  {Keski-Kuha}, R., {Kuchner}, M., {Lee}, R., {Lindler}, D., {Linfield}, R.,
  {Lisman}, D., {Lyon}, R., {Malhotra}, S., {Mathews}, G., {McCaughrean}, M.,
  {Mentzel}, J., {Mountain}, M., {NIkzad}, S., {O'Connell}, R., {Oey}, S.,
  {Padgett}, D., {Parvin}, B., {Procashka}, J., {Reeve}, W., {Reid}, I.~N.,
  {Rhoads}, J., {Roberge}, A., {Saif}, B., {Scowen}, P., {Seager}, S.,
  {Seigmund}, O., {Sembach}, K., {Shaklan}, S., {Shull}, M., and {Soummer}, R.
  (2009) ,
\newblock In {\em American Astronomical Society Meeting Abstracts}, Vol. 213 of
  {\em American Astronomical Society Meeting Abstracts}, p. 458.04

\bibitem[\protect\astroncite{{Stapelfeldt} et~al.}{2009}]{stapelfeldt_09}
{Stapelfeldt}, K.~R., {Krist}, J.~E., {Khan}, O., and {ATLAST Concept Study
  Team} (2009) ,
\newblock In {\em American Astronomical Society Meeting Abstracts}, Vol. 213 of
  {\em American Astronomical Society Meeting Abstracts}, p. 450.04

\bibitem[\protect\astroncite{{Th{\'e}bault} et~al.}{2008}]{thebault_08a}
{Th{\'e}bault}, P., {Marzari}, F., and {Scholl}, H. (2008) ,
\newblock {\em \mnras} {\bf 388}, 1528

\bibitem[\protect\astroncite{{Th{\'e}bault} et~al.}{2009}]{thebault_08b}
{Th{\'e}bault}, P., {Marzari}, F., and {Scholl}, H. (2009) ,
\newblock {\em \mnras} {\bf 393}, L21

\bibitem[\protect\astroncite{{Traub} et~al.}{2007}]{traub_07}
{Traub}, W., {Shaklan}, S., and {Lawson}, P. (2007) ,
\newblock In {\em In the Spirit of Bernard Lyot: The Direct Detection of
  Planets and Circumstellar Disks in the 21st Century}.  (P. {Kalas} ed.)

\bibitem[\protect\astroncite{{Trauger} et~al.}{2004}]{trauger_04}
{Trauger}, J.~T., {Burrows}, C., {Gordon}, B., {Green}, J.~J., {Lowman}, A.~E.,
  {Moody}, D., {Niessner}, A.~F., {Shi}, F., and {Wilson}, D. (2004) ,
\newblock In {\em Optical, Infrared, and Millimeter Space Telescopes. Edited by
  Mather, John C., Proceedings of the SPIE, Volume 5487, pp. 1330-1336 (2004).}

\bibitem[\protect\astroncite{{Trauger} et~al.}{2009}]{trauger_09}
{Trauger}, J.~T., {Stapelfeldt}, K., {Traub}, W., {Krist}, J., {Moody}, D.,
  {Serabyn}, E., {Mawet}, D., {Park}, P., {Henry}, C., {Gappinger}, R.,
  {Brugarolas}, P., {Dawson}, O., {Shaklan}, S., {Pueyo}, L., {Guyon}, O.,
  {Kasdin}, J., {Spergel}, D., {Vanderbei}, R., {Marcy}, G., {Brown}, R.~A.,
  {Schneider}, J., {Woodgate}, B., {Belikov}, R., {Matthews}, G., {Egerman},
  R., {Polidan}, R., {Lillie}, C., {Brady}, D., {Spittler}, C., {Ealey}, M.,
  and {Price}, T. (2009) ,
\newblock In {\em American Astronomical Society Meeting Abstracts}, Vol. 213 of
  {\em American Astronomical Society Meeting Abstracts}, p. 493.01

\bibitem[\protect\astroncite{{Trauger} and {Traub}}{2007}]{trauger_traub_07}
{Trauger}, J.~T. and {Traub}, W.~A. (2007) ,
\newblock {\em \nat} {\bf 446}, 771

\bibitem[\protect\astroncite{{Wade} and {Rucinski}}{1985}]{wade_85}
{Wade}, R.~A. and {Rucinski}, S.~M. (1985) ,
\newblock {\em \aaps} {\bf 60}, 471

\bibitem[\protect\astroncite{{Wertheimer} and {Laughlin}}{2006}]{wertheimer_06}
{Wertheimer}, J.~G. and {Laughlin}, G. (2006) ,
\newblock {\em \aj} {\bf 132}, 1995

\bibitem[\protect\astroncite{{Wiegert} and {Holman}}{1997}]{wiegert_1997}
{Wiegert}, P.~A. and {Holman}, M.~J. (1997) ,
\newblock {\em \aj} {\bf 113}, 1445

\end{thebibliography}

\end{document}